\documentclass[twocolumn,prl]{revtex4}
\usepackage{graphicx}
\usepackage{dcolumn}
\usepackage{bm}
\usepackage[dvipsnames]{color}

\begin{document}

\title{Bridging particle and wave sensitivity in a detector of 
configurable positive operator-valued measures}
\author{Graciana Puentes}
\email{g.puentes1@physics.ox.ac.uk}
\author{Jeff S. Lundeen}
\author{Matthijs P. A. Branderhorst}
\author{Hendrik B. Coldenstrodt-Ronge}
\author{Brian J. Smith}
\author{Ian A. Walmsley}
\affiliation{Clarendon Laboratory, University of Oxford, Parks Road 
OX1 3PU, United Kingdom}

\begin{abstract}
We report an optical detector with tunable positive operator-valued 
measures (POVMs). The device is based on a combination of weak-field 
homodyne techniques and photon-number-resolving detection.~The 
resulting POVMs can be continuously tuned from Fock-state projectors 
to a variety of phase-dependent quantum-state measurements by 
adjusting different system parameters such as local oscillator 
coupling, amplitude and phase, allowing thus not only detection but 
also preparation of exotic quantum states.~Experimental tomographic 
reconstructions of classical benchmark states are presented as a 
demonstration of the detector capabilities.
\end{abstract}

\maketitle

Detecting optical fields plays a key role throughout physics.~From the 
definition of the SI unit of luminosity \cite{Cheung}, to the 
characterization of quantum processes such as quantum logic gates 
\cite{qpt}, precise measurement of the electromagnetic field is 
central to both fundamental and applied physics.~The usual method of 
measuring fields in optics is either by counting photons or by 
measuring the amplitude and phase of the electric field. It is usually 
not possible to move continuously between these two measurement 
regimes with a single detector.~For example, standard homodyne 
detectors cannot directly probe the particle nature of light since it 
is masked by the local oscillator and electronic noise 
\cite{Raymer}.~Conversely photon-number-resolving (PNR) detectors 
possess no phase reference and thus have no sensitivity to the wave 
nature of light \cite{Achilles}.
 
\begin{figure}[tbp]
\includegraphics[width=0.50\textwidth]{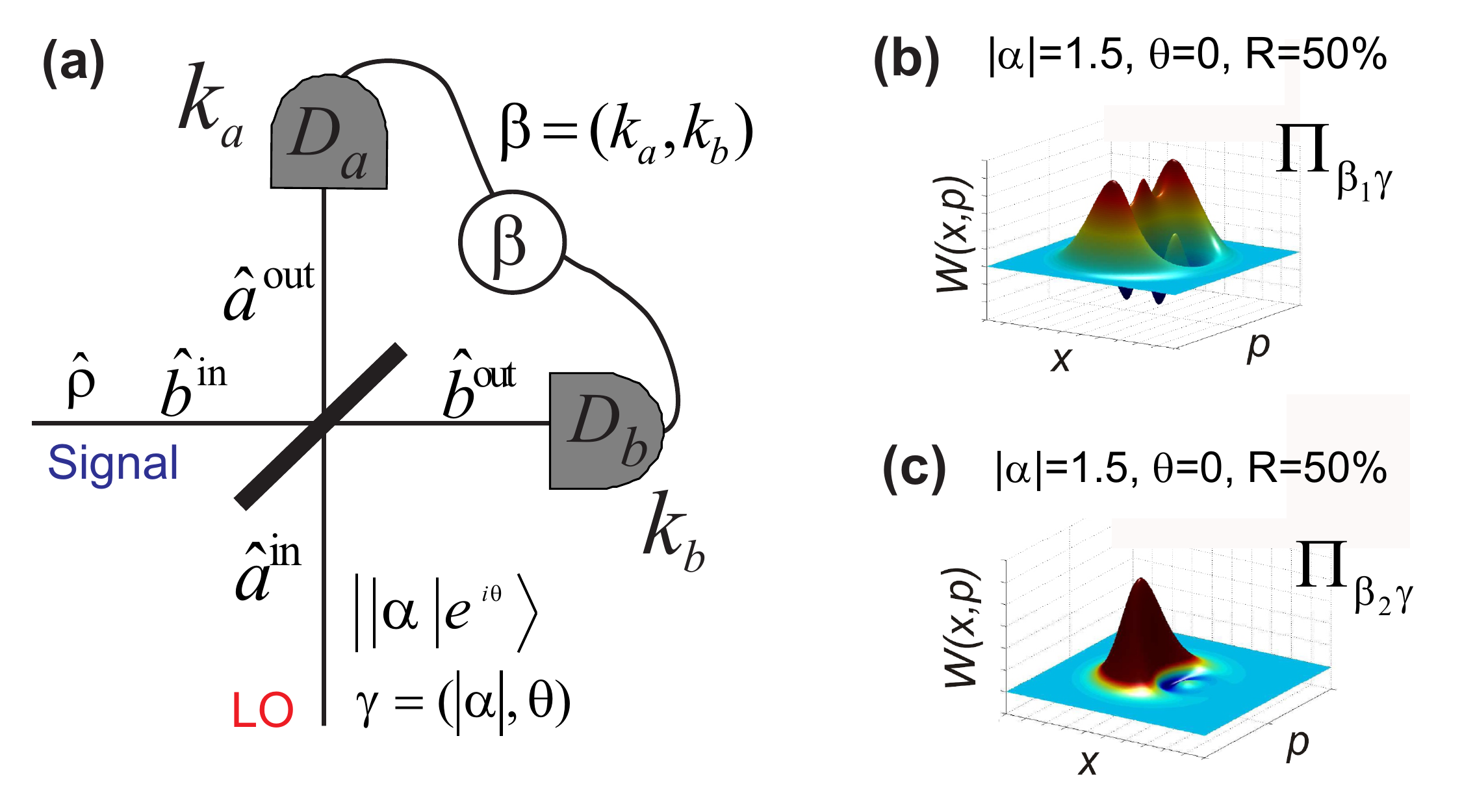}
\caption{(a)~Proposed scheme for the POVM configurable detector, $D_a$ 
and $D_b$ are PNR detectors, $\alpha$ is a weak coherent state; (b) 
and (c) POVM elements $\hat \Pi_{\beta \gamma}$ corresponding to click 
events (b) $\protect\beta_1 =(1,3)$ and (c) $\protect\beta_2=(1,1)$ 
for $LO$ settings $\gamma=(|\protect\alpha |=1.5,\theta=0)$, $R=50\%$ 
and $90\%$ efficient PNR detectors.~$\hat \Pi_{\beta \gamma}$ projects 
onto a single-mode (b) Schr\"{o}dinger-kitten state and (c) squeezed 
state, with high probabilities.}
\label{fig:1}
\end{figure}

The action of a given detector can be specified by a positive 
operator-valued measurement (POVM) set  $\{\hat \Pi_{\beta \gamma}\}$,  
where $\{\beta\}$ labels the outcomes and $\{\gamma\}$ labels the 
settings, such that for each $\gamma$ the set is complete, 
$\sum_{\beta} \hat \Pi_{\beta \gamma}=I$  \cite{Holevo}.~The 
probability of obtaining outcome $\beta$ for setting $\gamma$ and 
input state $\hat{\rho}$~is $p_{\beta\gamma}=\mathrm{Tr}[\hat 
\Pi_{\beta \gamma} \hat \rho]$.~In conventional detectors $\{\hat 
\Pi_{\beta \gamma}\}$ is fixed by the intrinsic nature of the 
device.~Typically, such POVMs
can encompass either Fock-state projectors (for PNR detectors) or
field-quadrature projectors (for homodyne devices). In this Letter, we
report a configurable detector with a flexible POVM, able to
transition smoothly from quadrature to photon-number detection.~The 
detector is constructed from a variable reflectivity ($R$) 
beam-splitter (BS), two PNR detectors, and an auxiliary weak coherent 
state acting as the local oscillator (LO). We denote such device as 
photon-number-resolving homodyne detector (PNRHD).~The BS input modes, 
labeled $\hat{a}^{\mathrm{in}}$ and $\hat{b}^{\mathrm{in}}$, 
correspond to the LO and the signal ($\hat \rho $), respectively (see 
Fig.~1~(a)). The output modes, labeled by $\hat{a}^{\mathrm{out}}$ and 
$\hat{b}^{\mathrm{out}}$, are detected by PNR detectors $D_{a}$ and 
$D_{b}$ giving joint outcomes $\{\beta=(k_{a},k_{b})\}$, where 
$k_{a(b)}$ labels the number of clicks registered at $D_{a(b)}$.~The 
adjustable local oscillator has settings $\{ \gamma=(|\alpha|,\theta) 
\}$, where $\alpha=|\alpha|e^{i\theta}$ represents the LO complex 
amplitude.~By tuning the LO coupling, amplitude and phase the detector 
POVM set can be configured to project onto a variety of fundamental 
quantum states of the radiation field -- 
Fock,~displaced-Fock,~quadrature-squeezed,~and Schr\"{o}dinger-kittens 
states for example.~Figures \ref{fig:1} (b) and (c) depict the Wigner 
representation \cite{Wigner} of two POVM elements ($\hat \Pi_{\beta 
\gamma}$), for two different detector outcomes $\{\beta_1,\beta_2\}$ 
and LO setting $\gamma$.~Here $\{\hat \Pi_{\beta \gamma}\}$ project 
onto (b) Schr\"{o}dinger-kitten states and (c) quadrature squeezed 
states with high probabilities. Since the action of a measurement is 
not only to reveal some property of the state of a system, but also to 
project the system in a state commensurate with that information, all 
measurement devices may in principle be used as preparation devices. 
Thus, the remarkable POVM elements of the PNRHD can be used not only 
to optimally detect the appropriate states \cite{LSE} but also to 
prepare them from quadrature-entangled beams \cite{Lvovsky}.   

As an experimental demonstration of the PNRHD capabilities, we have 
tomographically reconstructed a group of classical benchmark states. 
To our knowledge, this constitutes the first full tomographic 
reconstructions involving PNR detectors. Indeed, state tomography 
based on photon-counting detection has been the subject of much 
theoretical work \cite{Theory} and the few experimental 
implementations reported to date involve only binary (on/off) 
detectors \cite{Experiment}.~This is due, in part, to the relative 
infancy of PNR detector technology, which is an active area of 
research with several different approaches to photon counting 
\cite{PNRD}.~In the experiments presented here time-multiplexed PNR 
detectors are used, but the techniques can be readily extended to 
other PNR detectors.~The ability to change the measurement basis of 
the detector enables applications other than state reconstruction for 
this detection scheme, such as in non-local state preparation, 
precision quantum metrology \cite{Dowling}, or the implementation of a 
continuous-variable entanglement witness \cite{Horodecki}. 

The POVM elements are derived from an analytical model by  first 
considering  ideal PNR detectors, able to resolve $n$ photons (see 
Fig. \ref{fig:1} (a)). In this case the probability of obtaining 
measurement outcome $\beta=(n_{a},n_{b})$ for LO setting $\gamma$ is 
related to $\hat\rho$ by  \cite{Pregnell}  
\begin{equation}
\label{eq:1}
p_{\beta \gamma }=\mathrm{Tr}_{ab}[\hat{U}\hat{\sigma}_{ab} 
\hat{U}^{\dagger } |n_a\rangle \langle n_a|_a\otimes|n_b\rangle 
\langle n_b|_b],
\end{equation}%
\noindent with 
$\hat{U}=e^{i\xi(\hat{b}^{\dagger}\hat{a}+\hat{a}^{\dagger}\hat{b})}$ 
the unitary operator representing the BS, $R=\cos^2(\xi)$ the LO 
coupling, $\hat{\sigma}_{ab} =|\alpha \rangle \langle\alpha|_a \otimes 
\hat \rho_b$ the two-mode input state and $|n_{a(b)}\rangle$ the 
photon number states to be detected at $D_{a(b)}$.~Using cyclic 
properties of the trace,~Eq.~(\ref{eq:1}) can be written as $p_{\beta 
\gamma }=\mathrm{Tr}_{b}[\hat{\rho}_{b}\hat{\Pi}_{\beta \gamma}]$.~For 
the case of ideal detectors the POVM element $\hat{\Pi}_{\beta 
\gamma}^{\mathrm{ideal}}$ is a projector $\hat{\Pi}_{\beta 
\gamma}^{\mathrm{ideal}}=|\chi\rangle \langle \chi|_{b}$, where 
$|\chi\rangle_{b} = \langle 
\alpha_{a}|U^{\dagger}|n_{a}\rangle|n_{b}\rangle_b$. For $R=1/2$ this 
can  be expressed as
\begin{eqnarray}
\label{eq:3}
|\chi\rangle_{b} = e^\frac{-|\alpha|^2}{2}
\frac{(\alpha^{*}-i\hat{b}^{\dagger})^{n_a}(\hat{b}^{\dagger}-i\alpha^
{*})^{n_b}}{\sqrt{2^{(n_a+n_b)}n_a!n_b!}}|0\rangle_b.
\end{eqnarray}
In any realistic scheme one has to account for unavoidable 
imperfections in the PNR detectors such as loss, non-unit efficiency, 
and overlap in detector responses.~For a given PNR detector, this can 
be done by considering the detector design \cite{Achilles}, or by 
experimental detector tomography \cite{ColdenstrodtRonge2008}.~The 
time multiplexed detectors (TMDs) used in our experiments accept 
states of light contained in pulsed wave-packet modes.~Each incoming 
pulse is split into several spatial and temporal modes by a fiber beam 
splitter network that are subsequently registered by avalanche photo 
diodes (APDs).~Because the TMD alone is not phase sensitive, its 
operation can be described as a map from the incoming photon-number
distribution $\vec{\rho}$ (the diagonal components of the density 
matrix) to the measured click statistics $\vec{k}$ 
by $\vec{k} =\mathbf{C} \, \mathbf{L}\vec{\rho}$.~Here $\mathbf{L}$ 
and $\mathbf{C}$ are matrices accounting for loss and the intrinsic 
detector structure \cite{Achilles}, respectively. To calculate the 
POVM elements  implemented by our configurable PNR homodyne detector, 
the POVMs for TMD detectors $D_a$ and $D_b$ are determined from their 
$\mathbf{CL}$ matrices (characterized by independent methods).~The 
$n$-th element of the a(b) TMD POVM then replaces the projectors 
$|n_{a(b)}\rangle \langle n_{a(b)}|$ in Eq.~(\ref{eq:1}) to obtain the 
final expression for the POVM elements $\hat \Pi _{\beta \gamma }$.~We 
note that our TMDs can resolve up to 8 photons, setting the number of 
possible outcomes to 81 and truncating the operator Hilbert space to 
$9\times 9$ matrices.~

By adjusting the local oscillator amplitude and phase, the detector 
POVM elements $\hat \Pi_{\beta \gamma }$ can be tuned to project onto 
different bases. We illustrate this by calculating $\hat \Pi_{\beta 
\gamma }$ for two particular examples, shown in~Fig. \ref{fig:1}.~For 
a sufficiently large LO amplitude and detector efficiency 
($\eta$),~the joint click outcomes of the PNR detectors can nearly 
project onto a single-mode Schr\"{o}dinger-kitten state or  
quadrature-squeezed state, a feature which suggests our detector can 
be used both for preparation and direct detection of such quantum 
states.~Figure~\ref{fig:1} depicts the Wigner representation of two 
normalized POVM elements $\hat \Pi ^{\mathrm{norm}}_{\beta \gamma }$ 
for $\alpha=1.5$, $\eta_{a,b}=90\%$ \cite{NIST} and 50:50 beam 
splitter ratio, where $(x,p)$ label the phase-space conjugate 
variables.~Note that a normalized POVM element is a positive definite 
operator with unit trace, allowing it to have a phase-space 
representation similar to that of a density matrix.~To quantify the 
state preparation (detection) efficiency for these detector 
configurations, we calculate the overlap between the target state 
$\hat \rho^{\mathrm{tar}}$ (i.e., a Schr\"{o}dinger-cat-like state or 
squeezed state) and the normalized POVM projector $\hat \Pi 
^{\mathrm{norm}}_{\beta \gamma}$, by $p=\mathrm{Tr}[\hat \Pi 
^{\mathrm{norm}}_{\beta \gamma }\hat \rho^{\mathrm{tar}}]$.~In our 
numerical simulations, $p \approx 80\%$  can increase to values of up 
to $90\%$ for unit efficiency PNR detectors. 
\begin{figure}[tbp]
\includegraphics[width=0.40\textwidth]{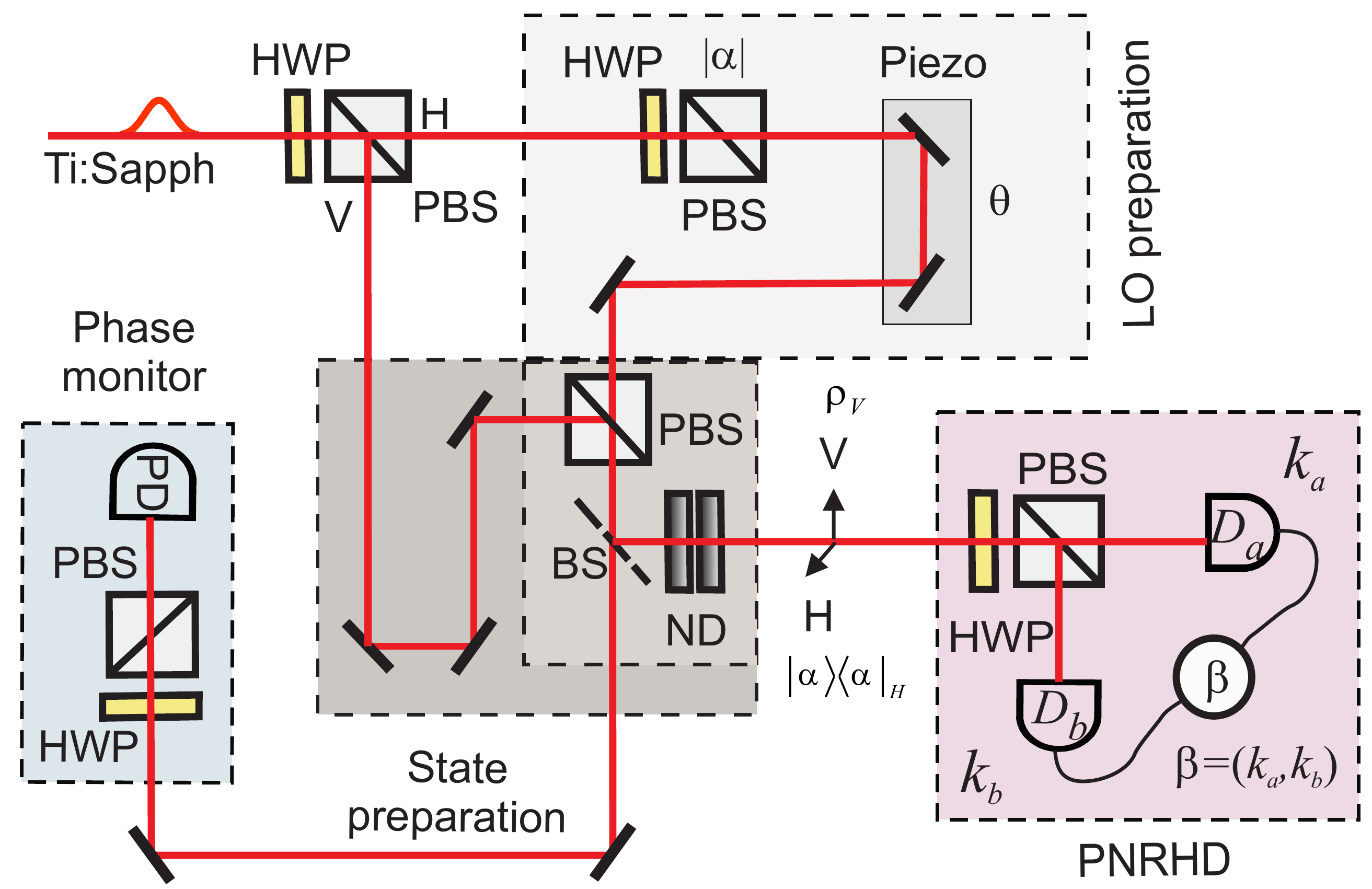}
\caption{Experimental set up for the photon number resolving homodyne 
detector (PNRHD).~The output of a Ti:Sapph laser is split into two 
arms (by a PBS), which eventually correspond to signal $\hat{\rho}$ 
and LO beams.~A piezo moves a mirror to set the phase  $\protect 
\theta$ of the LO.~Both arms are then recombined and sent through the 
configurable homodyne detector.}
\label{fig:3}
\end{figure}

To demonstrate the ability of a PNRHD we experimentally reconstruct 
various states derived from a coherent-state laser pulse using TMDs 
for the PNR detectors.~The experimental set up, shown in 
Fig.~\ref{fig:3}, consists of three main components: the state and 
local oscillator preparation, and the configurable PNRHD.~The signal 
and local oscillator are formed by splitting $90$~fs  Ti:Sapph laser 
pulses, centered at $784$~nm and cavity dumped to a repetition rate of 
$250$~kHz.~In the time-multiplexed detection scheme the APD dead-time 
($t_{D}\approx 50$~ns) in conjunction with the total number of 
temporal modes influences the maximal detection rate. We chose a time 
delay between temporal modes of $100$ ns after which the after-pulsing 
probability of the APDs drops below $0.1\%$. This, in combination with 
electronic time gating, makes after-pulsing effects negligible and 
reduces dark counts to less than $5$~cnt/sec. The input laser mode is 
split equally into two optical paths by a half-wave plate (HWP) and 
polarizing beam splitter (PBS), corresponding to $V$ and $H$ 
polarizations, respectively.~The LO path has a HWP and PBS to 
independently  control its amplitude, which is typically set to a 
small percentage of the total signal.~A pair of mirrors in the LO 
path, placed on a piezo-electric-controlled translation stage, set the 
relative phase ($\theta$) between  the LO and signal.~The two beams 
are recombined into a single path at a second PBS. A beam sampler (BS) 
with low reflectivity followed by a set of calibrated neutral density 
(ND) filters completes the state preparation and directs the signal 
and LO to a HWP-PBS combination that acts as the homodyne BS in 
Fig.~\ref{fig:1}~(a).~The outputs of the final PBS are sent to TMDs 
($D_{a,b}$).~The joint detection events are collected using a 
field-programmable gate array (FPGA) interfaced with a computer. The 
joint statistics are monitored for 100 different phase settings with a 
total measurement time of 15 minutes.~The relative phase $\theta$ can 
be monitored using the light transmitted through the BS,~which is then 
interfered using a HWP and PBS and detected at a photo diode (PD).~The 
typical fringe visibility measured at this phase-monitoring step, 
which indicates the mode match between the signal and LO, was 
approximately 70\%.~This relatively low contrast is due to chromatic 
dispersion in the beam paths and can be improved by placing 
narrow-band interference filters in the initial laser beam path. 
Material imperfections and  inaccurate calibration in the homodyne 
HWP-PBS  system could in principle be accounted for by the model  
parameter $R$, however such deviations were found to be below $1\%$ in 
our optical elements and are not considered a significant  source of 
experimental uncertainty in the error analysis.

The local oscillator amplitude $|\alpha|=\langle n_{\mathrm{LO}} 
\rangle^{1/2}$ is obtained by measuring the average photon number in 
the LO beam $\langle n^{\mathrm{meas}} \rangle$ and then multiplying 
by the transmission ($T$) of a set of  calibrated neutral density (ND) 
filters, so that $(T \langle n^{\mathrm{meas}} \rangle)^{1/2} = 
\langle n_{\mathrm{LO}} \rangle^{1/2}$, with relative error of 
approximately $5\%$, due to the inaccuracy in $T$.~The  detector 
efficiencies ($\eta_{a(b)}$) are obtained by fitting the parameters 
$\eta_{a(b)}$ in the loss matrix $\mathbf{L}$ which  retrieve an 
average photon number  $\langle n_{a(b)}\rangle \approx \langle 
n_{\mathrm{LO}} \rangle /2$,  measured at detector $D_{a(b)}$ when the 
signal is blocked. For our experiments the fitted detector 
efficiencies were $\eta_a=0.10 \pm 0.01$ and $\eta_b=0.15  \pm 0.02$, 
where the errors ($\varepsilon_{\eta} \approx 10\%$) are obtained by 
propagating the uncertainty in $\langle n_{\mathrm{LO}}\rangle$.~The 
low efficiency, as compared to the APD quantum efficiency ($\approx 60 
\%$), is due to the single-mode fiber network used to implement the 
PNR detection. To estimate the effect of these errors in the 
tomographic reconstructions, we built two POVMs $\{\Pi_{\beta 
\gamma}^{\mathrm{max(min)}}\}$~using the maximum~(minimum)~possible 
values around the mean for the set of parameters ($\langle 
n_{\mathrm{LO}}^{\mathrm{max(min)}} \rangle, 
\eta_{a}^{\mathrm{max(min)}}, \eta_{b}^{\mathrm{max(min)}}$).~We found 
that, while a change of up to $10\%$ in $\langle n_{\mathrm{LO}} 
\rangle$ does not affect the reconstruction (due to the relatively 
small amplitude of the LO, as compared to the signal), a change of 
$10\%$ in the efficiencies propagates into a relative error in the 
final estimated average photon number $\langle n^{\mathrm{est}} 
\rangle$ of $\approx 10\%$. 
\begin{figure}[tbp]
\includegraphics[width=0.45\textwidth]{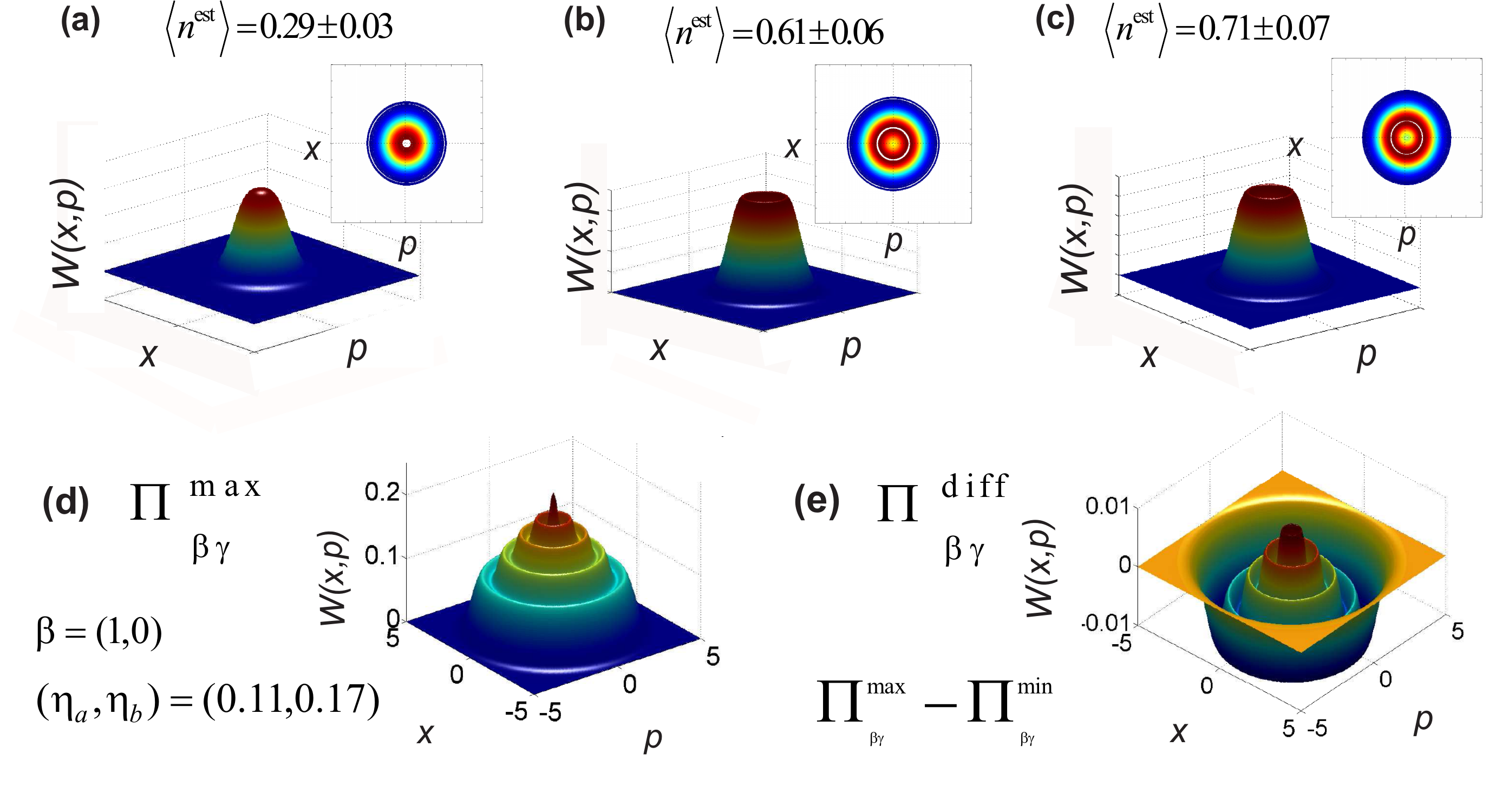}
\caption{Wigner function and corresponding contour plot (inset) of the 
experimentally reconstructed phase-averaged weak coherent states, with 
average photon numbers $\langle n^{\mathrm{est}} \rangle $ equal to 
(a) $0.29 \pm  0.03$, (b) $0.61 \pm 0.06$, and (c) $0.71 \pm 0.07$; 
(d) one extreme POVM element used in the error estimation 
$\hat{\Pi}^{\mathrm{max}}_{\beta\gamma }$, (e) difference between two 
extreme POVM elements  $\hat{\Pi}^{\mathrm{diff}}_{\beta\gamma 
}=\hat{\Pi}^{\mathrm{max}}_{\beta\gamma 
}-\hat{\Pi}^{\mathrm{min}}_{\beta\gamma }$.}
\label{fig:5}
\end{figure}
In order to estimate the most-likely state ($\hat 
\rho^{\mathrm{est}}$) that is compatible with the empirical 
photo-count statistics $p^{\mathrm{emp}}_{\beta\gamma}$, we use a 
recursive least-squares algorithm to minimize (over 
$\hat{\rho}^{\mathrm{est}}$) $\sum_{\beta 
\gamma}(p^{\mathrm{emp}}_{\beta \gamma}-\mathrm{Tr} [\hat{\Pi}_{\beta 
\gamma}\hat{\rho} ^{\mathrm{est}}])^2$, subject to the constraints 
$\hat \rho^{\mathrm{est}}\geq 0$ and $\mathrm{Tr} [\hat \rho 
^{\mathrm{est}}]=1$ \cite{LSE}. 

The first class of states to be examined were the phase-averaged 
coherent states.~Here 20 evenly-distributed LO phases $\theta $ are 
chosen between 0 and $2\pi$ for the data acquisition.~Figure 
\ref{fig:5} (a)-(c) show the experimentally reconstructed Wigner 
functions and corresponding contour plots for three different 
phase-averaged coherent states.~The Wigner functions are rotationaly 
symmetric and centered about the origin, as expected for such 
states.~To quantify the error in the estimated state we calculate the 
variance $\Delta$ between the two extreme estimated states 
$\hat{\rho}^{\mathrm{max(min)}}$, obtained using the two extreme POVMs 
$\{\Pi_{\beta \gamma}^{\mathrm{max(min)}}\}$, by  $\Delta=1-F$, with 
$F=|\mathrm{Tr}(\sqrt{(\sqrt{(\hat{\rho}^{\mathrm{min}})}
\hat{\rho}^{\mathrm{max}}\sqrt{(\hat{\rho}^{\mathrm{min}})} )})|^{2}$ 
the fidelity between the two extreme estimated density matrices. 
Figure \ref{fig:5} shows (d) an extreme POVM element for outcome 
$\beta=(1,0)$ and (e) difference between  
$\hat{\Pi}^{\mathrm{max}}_{\beta\gamma 
}-\hat{\Pi}^{\mathrm{min}}_{\beta\gamma }$. Such POVM elements do not 
look exactly like a projector onto a single photon state. This due to 
the low detector efficiency which mixes POVM elements corresponding to 
higher photon numbers.~Additionally, as a second figure of merit, we 
quantify the proximity of the reconstructed state 
$\hat{\rho}^{\mathrm{est}}$ with an ideal phase-averaged coherent 
state $\hat{\rho}^{PA}$ of average photon number $\langle 
n^{\mathrm{est}} \rangle$, using
the fidelity $F$, by replacing $\hat{\rho}^{\mathrm{max(min)}}$ by 
$\hat{\rho}^{\mathrm{PA(\mathrm{est})}}$.~The variances and 
fidelities~$(\Delta,F)$~resulted in (a) (0.004,0.985), (b) 
(0.010,0.973) and (c) (0.027,0.954), respectively. Note that the error  
increases with the average photon number, an expected effect which is 
due to the truncation of the Fock state space present in the modeled 
POVMs. 
\begin{figure}[tbp]
\includegraphics[width=0.35\textwidth]{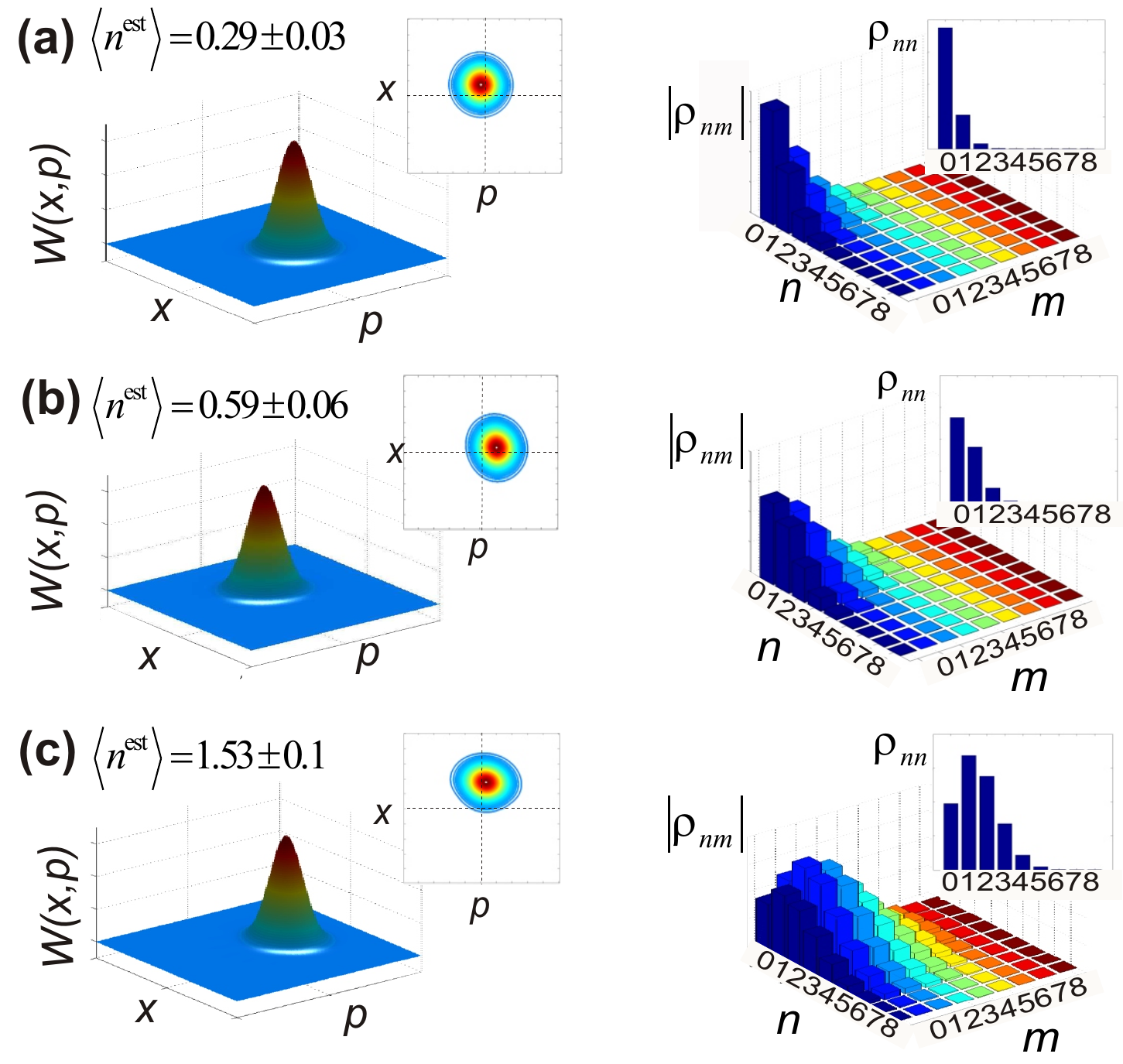}
\caption{Wigner function (left) and amplitude of the density matrix in 
the photon-number basis $|\hat \rho_{nm}|$ (right) for 
experimentally-reconstructed weak coherent states.~Here~$(x,p)$ labels 
the quadratures and $(n,m)$ labels the photon-numbers.~Insets show 
corresponding contour plot (left) and diagonal
matrix elements (right).~Rows correspond to an average photon number 
$\langle n^{\mathrm{est}}\rangle$ of (a) $0.29 \pm 0.03$, (b) $0.59 
\pm 0.06$, and (c) $1.53 \pm 0.1$.}
\label{fig:4}
\end{figure}

Next, in order to  test the phase sensitivity of this detector, 
several coherent states were tomographically reconstructed. Figure 
\ref{fig:4} shows the experimentally reconstructed Wigner function 
$W(x,p)$ and density-matrix amplitude $|\hat \rho |$, for coherent 
states of different average photon numbers. The variance and fidelity 
are used again as figures of merit of the state reconstruction, with 
$\hat \rho^{CS}$ the density matrix for an ideal coherent state with 
average photon number $\langle n^{\mathrm{est}}\rangle $ and phase 
$\theta^{\mathrm{est}}$ taking the place of $\hat \rho^{PA}$, $\langle 
n^{\mathrm{est}}\rangle$ was obtained from the diagonal elements of  
$\hat{\rho}^{\mathrm{est}}$ and
$\theta^{\mathrm{est}}$ was obtained as an average over the  
off-diagonal matrix elements using ~$\rho^{\mathrm{est}}_{n,m}=
\langle n^{\mathrm{est}}\rangle^{(n+m)/2}
e^{-(\langle n^{\mathrm{est}}\rangle)}
e^{-i(n-m)\theta^{\mathrm{est}}}/\sqrt{n!m!}$.~We 
found~$(\Delta,F)$~equal to (a)~(0.002,0.999), (b)~(0.010,0.998) and 
(c)~(0.010,0.997), respectively. 

In conclusion, we have introduced a highly adaptable homodyne 
detection scheme with PNR detectors.~This approach bridges the gap 
between phase-sensitive weak-field homodyne techniques and PNR 
detection, opening a new avenue of research and realm of applicability 
for PNR technology.~The various states the detector can directly 
project onto (squeezed states, Schr\"{o}dinger cats, and displaced 
Fock states) make it useful not only for state detection, but also for 
state preparation, a direction which is currently 
being explored at our group.~The detector has proven to be effective 
for state tomography as demonstrated by the experimental 
reconstruction of weak coherent and phase-averaged weak coherent 
states.

This work was supported by the EPSRC through the QIP IRC, by the EC 
under QAP, by the IST directorate, and by the Royal Society.~HCR is 
supported by the European Commission under the Marie Curie Program and 
by the Heinz-Durr Stipendienprogamm of the Studienstiftung des 
deutschen Volkes.


\begin{thebibliography}{99}

\bibitem{Cheung}Bureau International des Poids et Mesures, \emph{The 
international system of units (SI)}~(2006),~8th~ed., 
~www.bipm.org/en/si/si-brochure.

\bibitem{qpt}J. L. O'Brien \emph{et al}, Phys. Rev. Lett. \textbf{93}, 
080502 (2004).

\bibitem{Raymer}D. T. Smithey, M. Beck, M. G. Raymer, A. Faridani, 
Phys. Rev. Lett. \textbf{70}, 1244 (1993).

\bibitem{Achilles} D. Achilles \emph{et al}, J.
Mod. Opt. \textbf{51}, 1499 (2004).


\bibitem{Holevo} A. S. Holevo, \emph{Probabilistic and statistical 
aspects of quantum theory}~(North Holland, Amsterdam, 1982).

\bibitem{Wigner}W. P. Schleich, \emph{Quantum Optics in Phase 
Space}~(Wiley, Berlin, 2001).

\bibitem{LSE}R. L. Kosut, I. A. Walmsley and H. Rabitz, arXiv 
quant-ph/0411093 (2004).

\bibitem{Lvovsky}S. A. Babichev, B. Brezger, and A. I. Lvovsky, Phys. 
Rev. Lett. \textbf{92}, 047903 (2004).


\bibitem{Theory} G. Zambra and M. G. A. Paris, Phys. Rev. A 
\textbf{74}, 063830 (2006); M. Takeoka, M. Sasaki, and N. 
L\"{u}tkenhaus, Phys. Rev. Lett. 
\textbf{97} 040502 (2006); K. Banaszek, A. Dragan, K. W\'{o}dkiewicz, 
and C. Radzewicz, Phys. Rev. A \textbf{66}, 043803 (2002); L. S. 
Phillips, S. M. Barnett, and D. T. Pegg, Phys. Rev. A \textbf{58}, 
3259 (1998); D. Mogilevtsev, Z. Hradil, and J. Pe\v{r}ina, Quantum 
Semiclass. Opt. \textbf{10}, 345 (1998); S. Wallentowitz and W. Vogel, 
Phys. Rev. A \textbf{53}, 4528 (1996); K. Banaszek and K. 
W\'{o}dkiewicz, Phys. Rev. Lett. \textbf{76}, 4344 (1996).

\bibitem{Experiment}G. Zambra \emph{et al}, Phys. Rev. Lett. 
\textbf{95}, 063602 (2005); K. Banaszek, C. Radzewicz, and K. 
W\'{o}dkiewicz, and J. S. Krasi\'{n}ski, Phys. Rev. A \textbf{60}, 674 
(1999).

\bibitem{PNRD} A. Divochiy \emph{et al}, Nature Photonics \textbf{2}, 
302 (2008); B. E. Kardyna \emph{et al}, accepted in Nature Photonics
(2008).

\bibitem{Dowling} S. Huver, C. Wildfeuer, and J. Dowling,
arXiv/quant-ph/08050296 (2008).

\bibitem{Horodecki} M. Horodecki, P. Horodecki, and R. Horodecki, 
Phys. Lett. A \textbf{223}, 1 (1996).

\bibitem{Pregnell} K. L. Pregnell, and D. T. Pegg, Phys. Rev. A 
\textbf{66},
013810 (2002).

\bibitem{ColdenstrodtRonge2008} J. S. Lundeen \emph{et al}, Nature 
Physics \textbf{5}, 27 (2009).

\bibitem{NIST}An efficiency of $90\%$ is achievable with current PNR 
detectors (A. E. Lita, A. J. Miller, and S. W. Nam, Opt. Express 
\textbf{16}, 3032 (2008)).


\end{thebibliography}
\end{document}